\documentclass[11pt]{article}

\usepackage[margin=1in]{geometry}
\usepackage{amsmath,amssymb}
\usepackage{graphicx}
\usepackage{xcolor}
\usepackage[export]{adjustbox}
\usepackage{wrapfig}
\usepackage[percent]{overpic}
\usepackage{caption}

\usepackage{hyperref} 
\usepackage{parskip}  

\usepackage[backend=biber,style=numeric,sorting=none,sortcites=true]{biblatex}
\newcommand{\panel}[2]{%
  \begin{overpic}[width=\linewidth,height=157pt,keepaspectratio=false]{#2}%
    \put(2,94){\colorbox{white}{\footnotesize\bfseries (#1)}}%
  \end{overpic}%
}

\title{Magnetotransport Measurements on Single Polycrystalline Grains of Type-II Silicon Clathrate}

\author{%
  Sam Saiter*,
  Joseph P.~ Briggs,
  Yinan Liu,
  Audrey Faricy,
  Gavin Sher,
  Carolyn A.~Koh,\\
  Reuben T.~Collins,
  Meenakshi Singh*%
}

\begin{document}

\maketitle

\begin{center}
\small
\textbf{Affiliations}\\[4pt]

\textbf{Department of Physics, Colorado School of Mines, Golden, USA}\\
Sam Saiter; Audrey Faricy; Gavin Sher; Reuben T.~Collins; Meenakshi Singh\\[6pt]

\textbf{Department of Chemical and Biological Engineering, Colorado School of Mines, Golden, USA}\\
Joseph P. Briggs; Yinan Liu; Carolyn A.~Koh\\[8pt]

\textbf{Corresponding author(s)*:}\\
Sam Saiter* \ \texttt{ssaiter@mines.edu}\\
Meenakshi Singh* \ \texttt{msingh@mines.edu}
\end{center}
\begin{abstract}
The first low-temperature electronic transport characterization of individual polycrystalline grains of type-II silicon clathrate (Na$_x$Si$_{136}$, $x \ll 1$), isolated using microfabrication techniques, is reported. Structural characterization via Raman spectroscopy confirms that the isolated grains are largely devoid of amorphous silicon (a-Si). Temperature-dependent resistivity reveals multiple conduction regimes, including thermally activated freeze-out behavior and a transition to low-activation-energy transport at cryogenic temperatures, consistent with hopping conduction mechanisms. Hall measurements from 290~K to 3.5~K yield carrier concentration and mobility trends that correlate with the extracted activation energies, verifying $n$-type conduction. Additionally, gate-dependent conductivity measurements demonstrate electrostatic tunability at room temperature. Collectively, these results establish the magnetotransport parameters of single-grain type-II silicon clathrate and demonstrate its potential for future quantum and optoelectronic devices.
\end{abstract}
\pagebreak

\section{Introduction\label{Introduction}}

Silicon  clathrates are cage-like allotropes of silicon, existing as metastable alternatives to the traditional diamond cubic structure of silicon~\cite{beekman2015}. Solid-form clathrates are synthesized in the presence of alkali-metal guest atoms, most commonly Na, which occupy interstitial sites within the cage structure~\cite{kasper1965}. When the guest atom concentration is sufficiently low (Na$_x$Si$_{136}$ with $x\ll1$), silicon clathrates exhibit intrinsic semiconductor characteristics~\cite{kovnir2004}, marked by a wide, nearly direct band gap of approximately 1.9 eV~\cite{Liu2025}. This wide band gap, along with uniquely tunable electronic and optical properties, positions silicon clathrates as a promising candidate for various optoelectronic and energy applications, including photovoltaics, photodetectors, and thermoelectrics ~\cite{gryko2000,nolas2014review,8,13}. Recent electron paramagnetic resonance (EPR) studies of low-sodium (Na) type-II silicon clathrate films have also demonstrated relatively long spin lifetimes, suggesting potential for spintronic devices and quantum information processing applications~\cite{2,shankar2010}.  Clathrate devices would leverage the advantages of established Si-based fabrication infrastructure with potential for overcoming the limitations of diamond-Si, such as its indirect bandgap. However, while previous investigations have reported resistivity measurements performed on Si clathrate films and powders~\cite{gryko2000,moriguchi2000,saito1995,Stefanoski2010}, the present work targets phase-pure, film-grown Si clathrates. Comprehensive low temperature magnetotransport studies—including critical parameters such as Hall mobility and carrier concentration— do not exist for this material.  To realize the potential of silicon clathrates, device development strategies and knowledge of fundamental transport parameters are critical. These efforts are therefore the focal points of this study.

Type-II silicon clathrate (Si$_{136}$) can be grown as large single crystals under high Na partial pressure and slow Na deprivation of Na$_4$Si$_4$\,\cite{Stefanoski2011Simple,Stefanofski2012,6}, but achieving semiconducting behavior requires removing the vast majority of guest Na. In bulk, that de-sodiation is prohibitively slow~\cite{Slingsby2016NaDiffusion}. By synthesizing thin films, the time required to extract sufficient Na to reach the semiconducting regime is greatly reduced, and recent film studies demonstrate controllable reduction of Na toward the low-$x$ limit and associated electronic changes\,\cite{Yinan2021,Briggs2024APL,1,14}. A practical downside is the formation of a disordered/amorphous Si (a-Si) overlayer, which must be removed prior to transport measurements, as noted in prior thin film work~\cite{Yinan2021,Mott1973Clathrates}.

During thin film growth, type-II clathrate forms 10-50~$\mu$m clathrate polycrystalline grains (Fig.~\ref{fig:1}). Each polycrystalline grain is predominantly composed of various crystallographic orientations of Na$_x$Si$_{136}$. These polycrystalline grains are separated by 5-10~$\mu m$ deep trenches. The a-Si top-layer that blankets the clathrate material is typically removed via plasma etching. However, this does little to remove a-Si coating the trenches between polycrystalline grains. Remaining a-Si therefore forms a boundary interface between individual grains of clathrate material. Hence, probing the electrical transport of a film of many polycrystalline grains necessarily happens in series with a-Si material.

These amorphous boundaries influence measurements and complicate the determination of transport properties. For example, in one such early study, Mott theorized that the low-temperature conduction behavior, identified as Variable-Range Hopping (VRH), was attributed primarily to electron transport through disordered a-Si regions between polycrystalline grains, rather than within the crystalline clathrate framework itself~\cite{Mott1973Clathrates}. VRH is often observed in amorphous materials such as silica, arising from Anderson localization that causes band type conduction to rapidly diminish, even at higher temperatures~\cite{Ambegaokar1971}~\cite{Castner1999}. Single-grain measurements, therefore, aim to probe the characteristic transport of type-II clathrate, avoiding the influence of these amorphous intergranular layers. 

In this context, we report a novel methodological approach of isolating individual polycrystalline grains from many-grain thin-film samples, enabling explicit electrical and magnetotransport measurements. This work presents the first low temperature magnetotransport measurements of polycrystalline grains isolated in this way, and analyzes the temperature dependence of conductivity in type-II Si clathrate. By identifying and contacting individual polycrystalline grains via maskless photolithography, confirming their crystallinity through Raman mapping, and carrying out systematic low-temperature magnetotransport and gate-dependent conductivity measurements, we show both the feasibility of device integration and the characteristic electronic behavior of type-II silicon clathrates. These advances lay the groundwork for the development of clathrate-based quantum and optoelectronic devices.

\section{Experimental Methods \label{Experiment}}

\subsection{Single Grain Sample Preparation \label{Contacting methods}}

\begin{wrapfigure}{r}{0.5\textwidth}
\begin{center}
\includegraphics[width=0.48\textwidth]{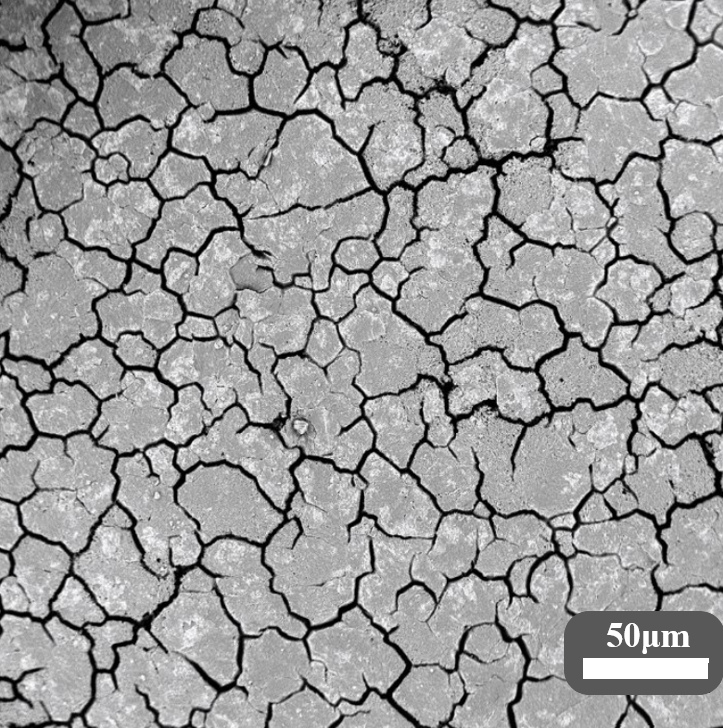}
\caption{\label{fig:1} Granular formation of type-II Si clathrate thin film. Inter-granular trenches are $\sim3-8~\mu m$ deep and consist of predominantly SiO$_2$.}
\end{center}
\end{wrapfigure}

Type-II Si Clathrates are grown on both p and n type lightly doped diamond Si substrates. The Si clathrate films used in this work were synthesized using a two-step process as discussed in detail elsewhere~\cite{Yinan2021}. These films average around 5~$\mu$m in thickness, and the trenches separating polycrystalline grains typically contain a-Si several microns deep (Fig. \ref{fig:1}). These trenches cause the film surface to be locally rough, and make depositing contiguous contacts between grains very difficult. Therefore, inter-granular transport measurements typically involve a-Si at the grain boundaries. To avoid these issues, we developed a process for transferring isolated grains to a different substrate,  then contacting them using optical lithography. 

A diamond silicon substrate with a 150~nm SiO$_2$ dielectric layer was coated with KMPR 1010, spun at 4500~rpm for 45s to form an 8~$\mu$m layer. KMPR is a thick photoresist often used for multilayer fabrication~\cite{15}, and was chosen for its very large electrical resistivity ($\sim10^{16}\Omega\cdot cm)$ and stability when exposed to various organic solvents common to lift-off processes~\cite{16}. After baking at 100$^\circ$C for 90 seconds, polycarbonate tweezers were used to scrape the film surface, loosening clathrate grains from the growth substrate and allowing them to be lightly brushed onto the KMPR coated substrate. A clean glass slide was then used to apply pressure to the clathrate grains, embedding them into the KMPR film. This secures the grains for further processing and provides partial planarization (Fig. \ref{fig:2}a). Despite not providing complete planarization, the KMPR layer afforded sufficient surface smoothing that, in conjunction with depositing contacts at a 45$^\circ$ angle, enabled continuous electrical coverage (Fig. \ref{fig:2}b). The added electrical isolation provided by the KMPR suppresses parasitic current leakage into the lightly doped Si substrate and reduces the likelihood of device failure from dielectric breakdown of the relatively thin SiO$_2$ layer. This is particularly important for low-temperature semiconductor measurements, where the impedance is often very large and correspondingly higher applied voltages are needed to obtain measurable signal levels.

Once polycrystalline clathrate grains were embedded in the KMPR, the sample was then returned to 100$^\circ$C for the remainder of its 5 minute pre-exposure bake, setting the film before the final cure. After curing the film by exposing in excess of 235 mJ/cm$^2$, it is subjected to a post-exposure bake of 2 minutes at 100$^\circ$C and a vigorous spraying of IPA. This spraying removes loose grains from the surface and helps isolate clathrate grains secured in the KMPR layer. The sample is then hard baked at 150$^\circ$C for 1 hour, as baking has been shown to increase the resistivity of KMPR to solvents~\cite{16}. Importantly, heating at any step of this fabrication process is limited, as above 350$^\circ$C we have observed the metastable silicon clathrate phase to begin to convert back to diamond silicon.

Once the polycrstalline clathrate grains are prepared as described above, standard photolithographic techniques can be used to contact the grains. S1813 Photoresist was spun on at 4500 rpm for 45 seconds, followed by a 60 second bake at 100$^\circ$C. Contacts were drawn using a Nanyte Beam maskless photolithography system, allowing for the tailoring of contacts to avoid unwanted grains in the surrounding KMPR, and for the particular geometry of a target grain. Samples were developed in MF319 developer for 30 seconds, followed by a 60s O$_2$ plasma clean at 100W with a gas pressure of 500 mTorr in an AutoGlow 200 Reactive Ion Etcher. An AJA UHV Multi-Technique Deposition system was used to deposit a 300~nm Al layer via magnetron sputtering followed by a 30~nm Au layer via electron beam evaporation. As mentioned, thick contacts deposited at a 45 degree angle are important for overcoming any roughness that remains around the edge of clathrate grains. The Au top layer was added for visibility in SEM (Fig. \ref{fig:2}b). 

Liftoff was completed via a 5 minute acetone soak followed by a brief ($<\!1$ second) dip into an acetone ultrasonic bath to remove remaining unwanted contact material. The sample was then ready for wire-bonding. Given the slight roughness of the contact pads resulting from small residual clathrate particulates that covered the KMPR layer, in addition to the sensitivity of the sample to heat, indium spheres were placed on the gold ball of a manual ball bonder (TPThb05). This increased adhesion between the gold ball and contact pads, enabling room temperature bonding.  

\subsection{Characterization Techniques \label{single}}

\begin{figure*}[t]
\centering
\begin{minipage}[t]{0.32\textwidth}
  \panel{2a}{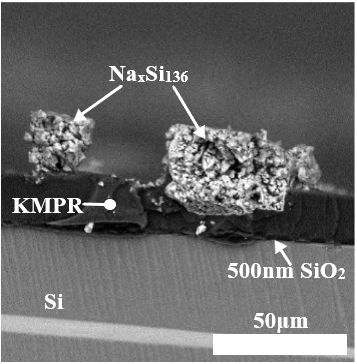}\\[6pt]
  \panel{2f}{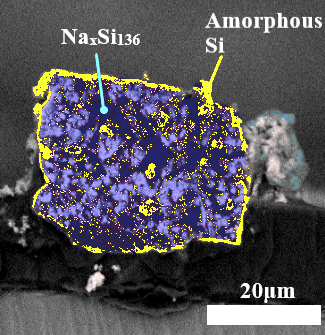}
\end{minipage}\hfill
\begin{minipage}[t]{0.32\textwidth}
  \panel{2b}{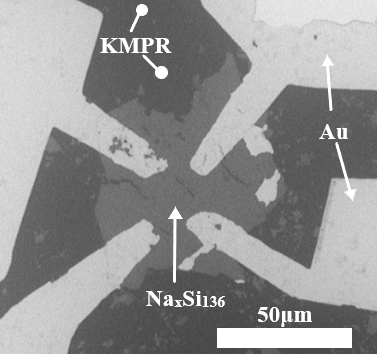}\\[6pt]
  \panel{2e}{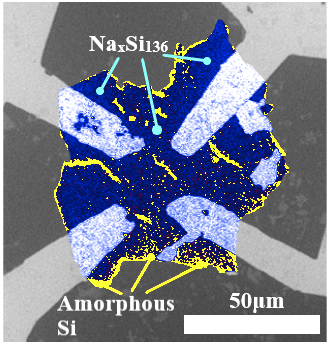}
  
\end{minipage}\hfill
\begin{minipage}[t]{0.32\textwidth}
  \panel{2c}{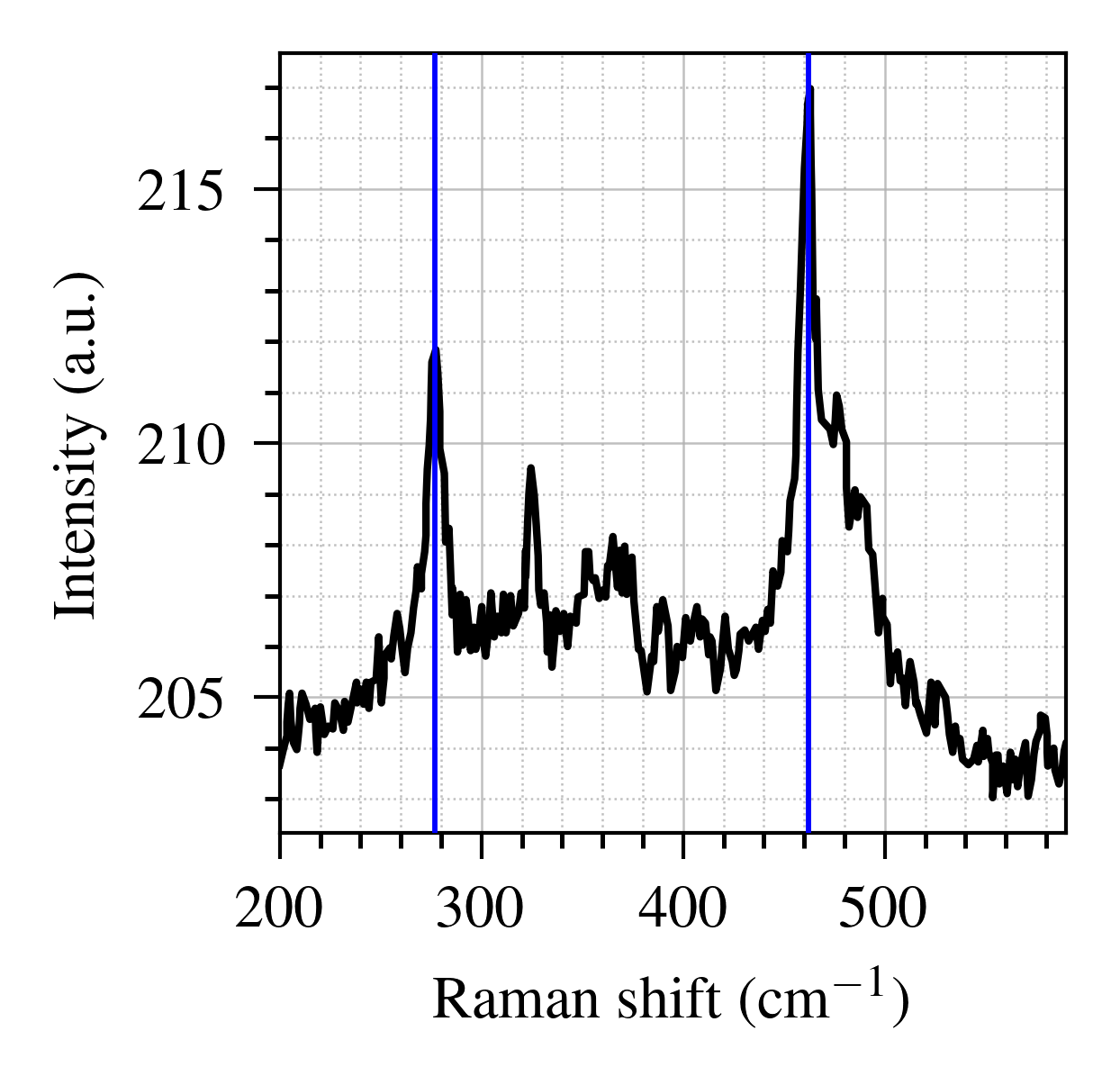}\\[6pt]
  \panel{2d}{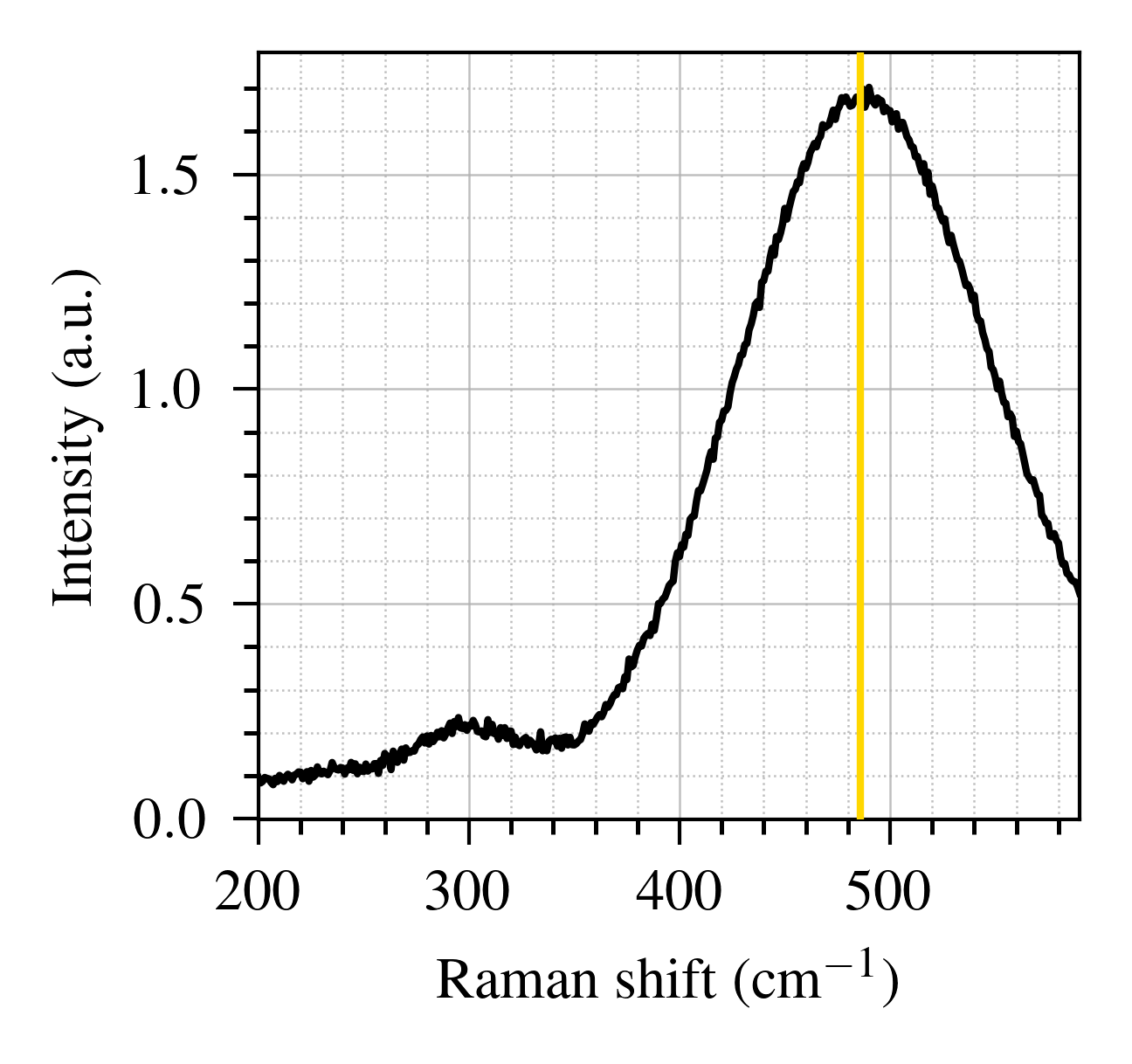}
\end{minipage}
\caption{Device Geometry and Raman Overview. (a) Cross section of prepared single-grain embedded in a layer of KMPR photoresist. (b) Top view of the devices with patterned Au contacts. (c) Raman spectrum of a pure polycrystalline Na$_x$Si$_{136}$ grain, with characteristic peaks identified ($\sim\!270$ and $\sim\!460$). (d) a-Si spectrum. (e) Top-view Raman map of polycrystalline clathrate grain with the same parameters as 2d. (f) Detailed Raman scan of single grain cross section. Blue areas have dominant clathrate peaks (as in c), and yellow indicates an a-Si peak (as in d). }
\label{fig:2}
\end{figure*}

A WITec Alpha300 Raman spectrometer was used to identify amorphous regions on the top of polycrystalline clathrate grains. This was used as a test to ensure successful removal of a-Si via plasma etching. Fig. \ref{fig:2}c shows one such spectrum measured, with identified peaks indicative of clathrate material and devoid of the a-Si peaks measured between grains. Similarly, \ref{fig:2}d shows a spectrum of the broad a-Si peak prevalent in the inter-granular trenches. A KMPR sample prepared with embedded polycrystalline clathrate grains was also cleaved, and the cross-section of a single grain located. This cross section was mapped by a high resolution Raman scan performed with a Tescan S8252G Raman SEM/FIB (Fig. \ref{fig:2}e and \ref{fig:2}f), the interpretation of which is discussed in the following section. Raman maps were converted to false color phase maps by baseline correcting each spectrum and integrating the clathrate bands near 270 and 460~cm$^{-1}$ to obtain $A_{270}$ and $A_{460}$. These normalized metrics $A$ are representative of the intensity of a particular peak, and a similar integration was performed for the amorphous Si peak to obtain $A_a$. A clathrate fraction was then computed as $X_{\mathrm{cl}}=\frac{A_{270}+A_{460}}{(A_{270}+A_{460})+A_a}$ and mapped to blue and yellow using threshold in $X_{\mathrm{cl}}$, with blue corresponding to clathrate rich regions and yellow corresponding to amorphous rich regions. Darker blue regions indicate higher apparent clathrate peak intensity, although we attribute much of this contrast to local variations in collection efficiency caused by surface roughness and angle dependent reflection rather than true changes in composition. Finally, these maps were overlayed on SEM micrographs of the polycrystalline clathrate grains.
   
Low-temperature transport measurements were performed using an AttoDRY2100 closed-cycle helium cryostat equipped with a 12T electromagnet. Measurements in a rough van der Pauw four-corner configuration were made to calculate the resistivity of the polycrystalline clathrate grains. Despite being pressed into a KMPR isolation layer, the boundary between the edges of clathrate grains and the KMPR sublayer spanned by Au contacts remained locally rough relative to typical photolithographic samples. This often produced asymmetries in the contacts as some residual Au resisted lift-off despite aggressive solvents and ultrasonication. Therefore, current reversal and field reversal were very important for solving for the geometric parameters of the Van der Pauw equation. 

Current was sourced using a Keithley 6221 AC and DC current source, and voltage was measured using a Keithley 2401 source-meter. As a result of carrier freeze out in the Si clathrate at low temperatures, these measurements were necessarily high impedance. Due to the large voltages and small currents intrinsic to high impedance measurements, all electronics shared a chassis ground direct to earth, and guarded connections were made to reduce parasitic effects. An excitation current of 100nA was applied in order to maximize the voltage signal while staying within the 200V maximum source and 200V maximum measurement input. DC leakage current for the system was measured to be $<\!10$pA when sourcing 200V on an open circuit, and when measuring a thick-film $100 \mathrm{G}\Omega$ test resistor, deviation in expected resistivity was measured to be $<\!1\%$ for the full range of measurement. 

\section{Results and Discussion\label{Results}}

\subsection{Materials characterization}
In order to confirm the removal of a-Si from our electrical characterization, Raman spectroscopy was utilized. An initial Raman spectrum collected by measuring on the top of a single polycrystalline grain (Fig. 2c) exhibits the clathrate-characteristic modes near $\sim\!270$ and $\sim\!460\,\mathrm{cm}^{-1}$ and is notably devoid of the broad amorphous Si band at $\sim\!480$–$500\,\mathrm{cm}^{-1}$. In contrast, spectra taken on amorphized intergranular regions show a dominant $\sim\!480$–$500\,\mathrm{cm}^{-1}$ feature (Fig.~2b). These peaks were used as a baseline to determine the relative quantity of a-Si and clathrate material in the larger, high resolution Raman scans performed on the top and cross section of the measured polycrystalline clathrate grain (Fig. \ref{fig:2}e and \ref{fig:2}f). The resulting maps indicate that most a-Si is confined to a thin exterior region surrounding the grains, and scans across multiple grains exhibit the same behavior. 

Additionally, clusters of a-Si inclusions are small and somewhat isolated relative to the amount of clathrate material within the internal structure of the polycrystalline grain. Because a-Si is substantially more resistive than crystalline clathrate, the combination of (i) spectral purity at the grain top after etch (Fig.~2e), (ii) low a-Si fraction in cross section (Fig.~2f), and (iii) consistency across many grains implies that current paths exist predominantly through the crystalline clathrate framework rather than through amorphous regions. During testing, this information was used to assert that a Raman scan of the top of a particular grain is sufficient to determine that clathrate current paths dominate, and it establishes that the electrical measurements of this work probe pure Na$_x$Si$_{136}$ intragranular transport with high confidence.

\subsection{Temperature-Dependent Resistivity}

Resistance was measured from 3 to 300~K (Fig.~3a). After averaging over all van der Pauw configurations, we plot $\ln(\rho)$ versus $1000/T$ (Fig.~3b) and identify three transport regimes with distinct Arrhenius slopes: intrinsic conduction ($T>200$~K), donor freeze-out ($60<T<200$~K), and low-temperature hopping ($T<60$~K). From the linear Arrhenius region in the freeze-out regime ($60<T<200$~K), we extract an activation energy of $\epsilon_{1}=65$~meV, while the low-temperature hopping regime ($T<60$~K) yields $\epsilon_{3}=0.3$~meV.

Cros \textit{et al.} reported the intrinsic $\rightarrow \epsilon_{1}$ crossover in Na$_x$Si$_{136}$ as $x$ decreased from 11 to 3, with the transition temperature shifting from $\sim$100~K to $\sim$150~K. Consistent with this trend, we observe the intrinsic $\rightarrow \epsilon_{1}$ transition at $\sim$270~K for $x \ll 1$. Gryko \textit{et al.} likewise observed an intrinsic $\rightarrow \epsilon_{1}$ crossover. In contrast to these prior studies, our measurements extend to sufficiently low temperature to resolve the subsequent $\epsilon_{1}\rightarrow\epsilon_{3}$ crossover into the hopping regime, which occurs near 60~K.

The extracted value of $\epsilon_{1}$ is consistent with the expected ionization energy of Na donors in Si~\cite{Doubrava1970}. Interestingly, prior measurements from our group indicated substantially shallower states ($\sim$19~meV) from photoluminescence. We attribute this discrepancy to the additional temperature dependence of carrier scattering inherent to the optical probe, which is discussed further below. Overall, $\epsilon_{1}=65$~meV is consistent with shallow donor states in Na-doped Si.

\begin{wrapfigure}{l}{0.5\textwidth}
\begin{center}
\includegraphics[width=0.48\textwidth]{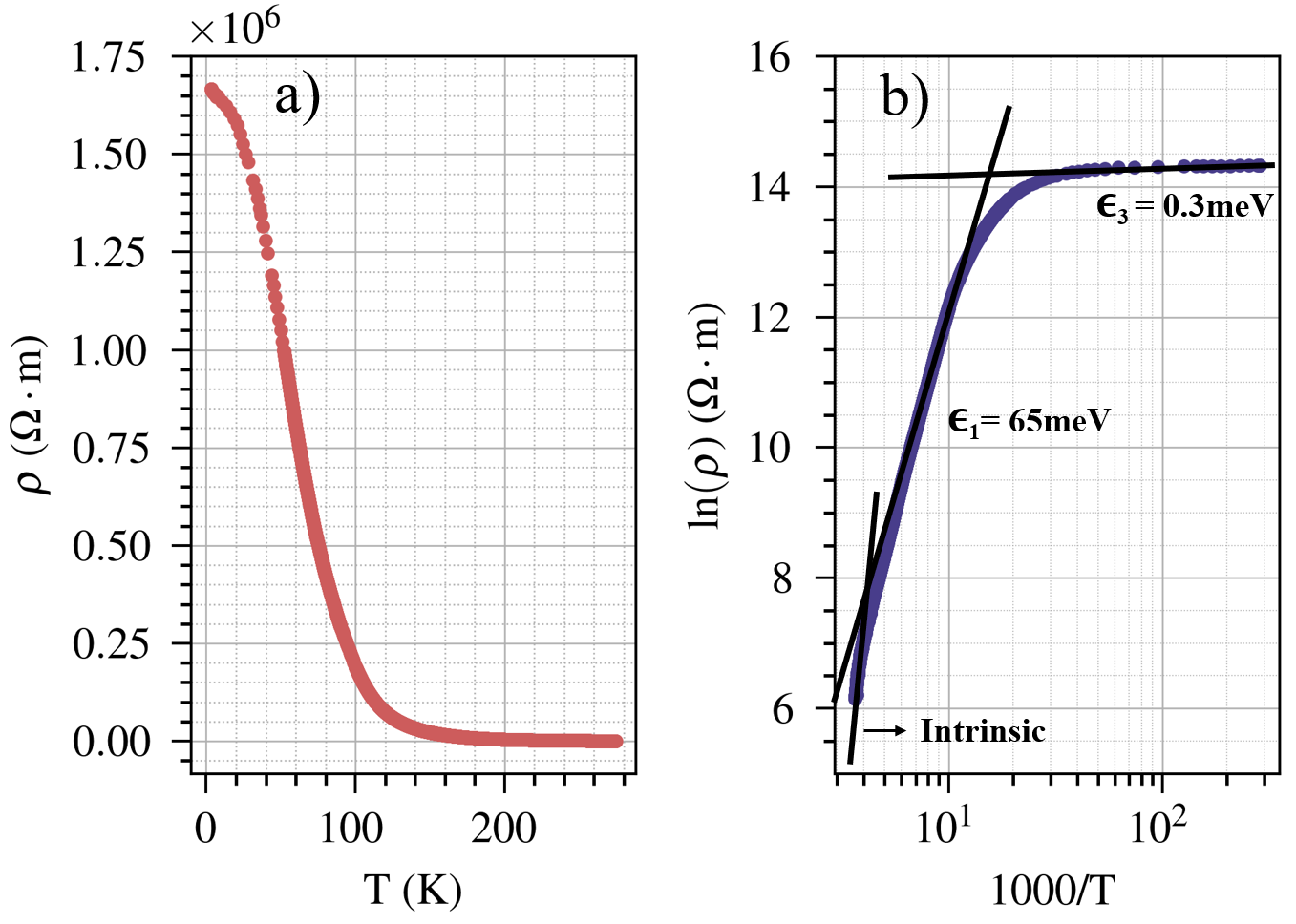}
\caption{\label{fig:rho_vs_T}(a) Temperature dependence of resistivity of single polycrystalline clathrate grain in the range of 3-300~K. (b) Arrhenius representation of the resistivity, $\ln(\rho)$ versus $1000/T$, highlighting three transport regimes and the corresponding low-temperature activated fits. Intrinsic conduction dominates from 290 to 200~K. Between $\sim$60 and 200~K, transport is governed by donor freeze-out with activation energy $\epsilon_1$. At the lowest temperatures (3--15~K), the data are consistent with a hopping regime characterized by an activation energy $\epsilon_3$.}
\end{center}
\end{wrapfigure}

The extracted value of $\epsilon_{3}=0.3$~meV is smaller than the $\epsilon_{3}=0.85$~meV reported previously for Na-doped silicon~\cite{Doubrava1970}. This reduction may arise from the degree of compensation, which is  consistent with earlier impurity-conduction studies~\cite{Ray1961}. Alternatively, the unusually small $\epsilon_{3}$ may indicate the onset of VRH.

To assess whether the low-temperature behavior is better interpreted as simple activated hopping or the onset of VRH, we now briefly outline the impurity-band transport framework and its common ambiguities near the insulating side of the metal--insulator transition (MIT)~\cite{ZabrodskiiZinovyeva1984}. Temperature-dependent impurity-band and localization physics make these kinds of low-temperature conductivity measurements difficult to interpret. Even for n-type diamond-cubic Si near the transition, the dominant transport picture remains under discussion~\cite{Danilyuk2017}. In many semiconductors, freeze-out conduction ($\epsilon_1$) gives way to various types of hopping conduction as temperature decreases. Hopping conduction---whether nearest-neighbor hopping (NNH) or variable-range hopping (VRH)---can be further complicated by weak localization and electron--electron interactions. These overlapping transport regimes make extracting definitive parameters such as activation energies or localization lengths difficult, particularly in materials that become highly resistive over the low temperature ranges of interest. 

With this in mind, we now outline a standard framework for interpreting temperature-dependent resistivity in lightly doped crystalline semiconductors. At sufficiently high temperatures, transport is dominated by thermally excited intrinsic carriers (intrinsic conduction). As the temperature decreases, phonon-assisted band transport diminishes and carriers are increasingly recaptured by donor states, leading to a steep drop in the free-carrier concentration. We refer to this as the freeze-out regime ($\epsilon_1$), in which the free-carrier concentration $n$ decreases approximately exponentially with decreasing temperature. In this region transport is dominated by the phonon-assisted activation of donor carriers. Accordingly, the activation energy $\epsilon_1$ is closely related to the donor ionization energy. At still lower temperature, donor carriers also become strongly localized, producing a further increase in resistance characterized here by an activation energy $\epsilon_3$. We refer to this as the hopping regime.
 
In disordered or lightly doped semiconductors, Anderson localization suppresses the spatial spreading of electronic wavefunctions. The impurity band---the aggregate of states introduced by donors---can therefore be modeled as a set of localized states of the form
\begin{equation}
\psi_i(\mathbf{r}) \sim \frac{1}{\xi^{3/2}} \exp\!\left(-\frac{|\mathbf{r}-\mathbf{R}_i|}{\xi}\right),
\end{equation}
where $\xi$ is the localization length and $\mathbf{R}_i$ the impurity position~\cite{Mott1973Clathrates}. These states weakly overlap between impurity sites, allowing carriers to tunnel-- or `hop'-- from site to site. The magnitude of $\epsilon_3$ is determined by a variety of factors, including the degree of wavefunction overlap arising from the distance between donor sites, the level of disorder in the spatial distribution of these sites, the distribution of states within the impurity band, and the level of compensation in the material~\cite{ShklovskiiEfros1984}. 

Therefore, for many lightly doped ($N_\mathrm{D} < 10^{17}\mathrm{cm}^{-3}$) crystalline semiconductors, the expression
\begin{equation}
    \rho(T) = A_{1} e^{\epsilon_{1}/k_{B}T} + A_{3} e^{\epsilon_{3}/k_{B}T},
\end{equation} 
is sufficient to accurately describe the low temperature electrical transport, where $A_{1}$ reflects the resistivity prefactor for donor ionization (linked to carrier concentration and effective mass) and $A_{3}$ corresponds to the hopping prefactor (sensitive to localization length and disorder). Because of their strong dependence on important physical parameters such as doping level, compensation, and level of disorder~\cite{Miller1960},\cite{Poklonski2025}, characterizing $\epsilon_1$ and $\epsilon_3$ is a common approach for probing lightly doped semiconductors. Indeed, assuming simple activated hopping transport, the analysis performed in Fig. would be sufficient to completely describe transport in type-II Si clathrate. However, the exact character of transport in the hopping regime varies from semiconductor to semiconductor ~\cite{15},\cite{16}. This is because the probability of hopping between states separated by spatial distance $R_{ij} = |\mathbf{R}_i - \mathbf{R}_j|$ contains both spatial and energetic contributions:
\begin{equation}
\Gamma_{ij} \;\propto\; \exp\!\left(-\frac{2R_{ij}}{\xi}\right)\,
\exp\!\left(-\frac{\Delta E}{k_B T}\right),
\end{equation}
where $\Delta E$ is the energy difference between sites, $\xi$ is the localization length, $k_B$ is Boltzmann's constant, and $T$ is temperature. 
For higher temperatures within the hopping transport regime, carriers minimize $\Delta E$, leading primarily to NNH. 
As temperature decreases, it can become favorable to accept a longer hopping distance $R_{ij}$ in order to access a site with a smaller energetic mismatch $\Delta E$. This competition between spatial and energetic factors gives rise to the VRH regime.

Mott showed that optimizing this balance yields a resistivity law that can be written compactly as~\cite{Mott1979}:
\begin{equation}
    \rho(T)=\rho_0\exp[(T_0/T)^p].
\end{equation}
where the characteristic exponent $p$ can be directly connected to different types of transport. This picture can be even further complicated by other physics within the material, such as stronger Coulomb interactions that influence the density of states in the impurity band, as in the case of Efros--Shklovskii variable-range hopping (ES-VRH). Various hopping phenomena have been reported in a wide variety of materials, either in place of NNH or as a transition at yet lower temperatures~\cite{15},\cite{16}. Generally, the characteristic exponent is taken to be $p=1$, $1/2$, and $1/4$ for NNH, Mott VRH, and Efros--Shklovskii VRH, respectively. Accurate determination of $p$ is thus critical to a clear picture of electrical transport in clathrates at the lowest temperatures. 

Classic work by Mott~\cite{Mott1973Clathrates} suggested that clathrate resistivity could follow a $T^{-1/4}$ VRH dependence across a broad range. As Mott suggests, these results are consistent with measurements contaminated by a-Si, as VRH has been observed in a-Si up to 300K~\cite{vanHapert2002}. Given the substantially lower a-Si content in the present devices, any VRH crossover would be expected to occur at lower temperature, if present. However, a simple Arrhenius fit is typically insufficient for determining VRH behavior, as many curves appear linear when plotted against $T^{-1/4}$. This ambiguity prompted Zabrodskii \textit{et al.}~\cite{Zabrodskii1981} to develop an analysis that distinguishes VRH from simple activated transport. By defining $W(T)=d\ln\sigma/d\ln T$, where $\sigma$ is the conductivity, and plotting $\ln W$ versus $\ln T$, the slope yields the hopping exponent $p$. This parameter can be used to directly identify the conduction mechanism. Using this method, we find that VRH-like behavior emerges only below $\sim$10~K. However, the high noise floor at these extreme resistances (G$\Omega$) prevents an unambiguous determination of whether the transport in this region corresponds to NNH or VRH.
 
Thus, while $\epsilon_{3}$ is well described by an activated form, its unusually small value and the Zabrodskii analysis suggest a crossover to VRH may be occurring. Indeed, comparison with the relatively high temperature ($>50$K) VRH previously reported~\cite{Mott1973Clathrates} suggests that a decrease in disorder by eliminating a-Si at grain boundaries may be responsible a reduced VRH crossover temperature. Low-temperature VRH could therefore be a natural consequence of reducing amorphous material in our clathrate devices. Regardless of whether the lowest-temperature regime is best described as activated hopping ($\epsilon_3$) or VRH, the observation of a below-freeze-out crossover is, to our knowledge, the first direct measurement of the $\epsilon_{1}\rightarrow\epsilon_{3}$ transition in type-II clathrates.

\subsection{Hall Measurements}
\begin{wrapfigure}{r}{0.5\textwidth}
\begin{center}
\includegraphics[width=0.48\textwidth]{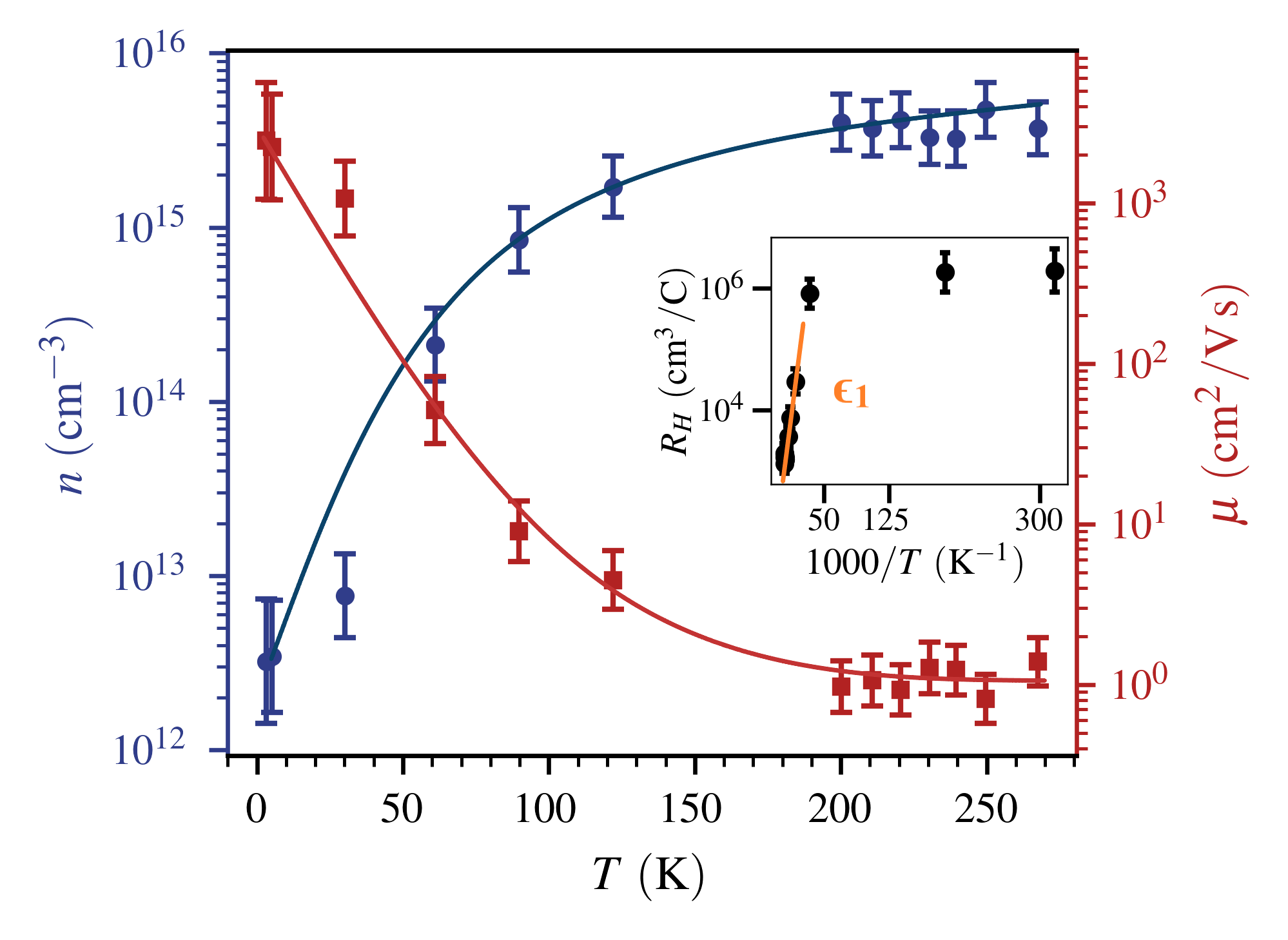}
\caption{\label{HallResults}Extracted Hall carrier concentration $n_H(T)$ and Hall mobility $\mu_H(T)$ for single-grain type-II silicon clathrate. Inset: $\ln|R_H|$ versus $1000/T$, highlighting the linear freeze-out region used to fit $\epsilon_1$ and the low-temperature plateau ($T \lesssim 20$~K) associated with the onset of hopping transport.}
\end{center}
\end{wrapfigure}
The temperature dependence of the Hall carrier concentration $n$ and Hall mobility $\mu$ were calculated from Hall voltages measured on a single polycrystalline grain of silicon clathrate at different temperatures (Fig. \ref{HallResults}). For a single dominant carrier type, the carrier density follows directly from the Hall coefficient ($R_H$), while the Hall mobility is obtained from $\mu = |R_H| \sigma$. The calculated carrier concentration $n(T)$ decreases by nearly three orders of magnitude between 100~K and 10~K, consistent with thermal activation from Na donor states.The onset of this decrease occurs near 200~K—remarkably close to the transition temperature extracted from our resistivity data for $\epsilon_1$ (65~meV). This correspondence supports our identification of a freeze-out regime below this temperature, in line with the donor activation behavior previously observed in clathrate materials~\cite{Cros1970,gryko2000}.
The temperature dependence of $n$ closely tracks the canonical behavior reported for lightly doped $n$-Si by Morin and Maita~\cite{Morin1954}. In particular, our $n(T)$ is comparable to their As-doped samples with $n\approx 1.75\times 10^{16}\,\mathrm{cm}^{-3}$: at $T\approx 250$~K, both datasets yield $n\sim 10^{16}\,\mathrm{cm}^{-3}$. Below $\sim$20~K, we observe a plateau in $R_H$ (inset, Fig.~\ref{HallResults}). In lightly doped semiconductors, this suggests a transition to hopping transport, as $R_H(T)$ typically peaks and then flattens as extended-state transport gives way to hopping~\cite{Friedman1981,Avdonin2016}. As an internal check, we fit the measured $n(T)$ to extract $\epsilon_1$, using the expected activated form over the freeze-out regime,
$n(T)\propto \exp\!\left(-\frac{\epsilon_1}{k_B T}\right)$. 
Propagating the uncertainty in $R_H$ into $n_H(T)$ yields a  range of acceptable $\epsilon_1$ values, within which $\epsilon_1=65$~meV is consistent. This is consistent with the value obtained from resistivity measurements and with the energy scale discussed by Morin and Maita~\cite{Morin1954}. Interestingly, however, a best fit gives a value $\epsilon_1 = 19$~meV. This is consistent with the $\sim$19~meV value obtained from temperature-dependent photoluminescence by Liu~\cite{Yinan2021}, supporting the interpretation that differences in apparent $\epsilon_1$ largely reflect the temperature dependence of the mobility, $\mu(T)$.

Type-II silicon clathrate (Si$_{136}$) exhibits a markedly different phonon dispersion from diamond-structured Si~\cite{Dong1999}. Its three acoustic branches are significantly flatter, reaching only $\sim$200~cm$^{-1}$ at the Brillouin-zone boundaries—about half the maximum frequency in diamond Si—and they overlap with numerous low-lying optical modes. In contrast, diamond-cubic Si has steep acoustic branches dispersing up to $\sim$400 cm$^{-1}$ with a wide phonon gap preceding its optic modes. The downward shift and flattening of the clathrate phonon bands indicate weaker average Si–Si bonding and a high density of low-frequency vibrational states, which enhance acoustic-phonon scattering and thereby suppress charge-carrier mobility. Nevertheless, the exceptionally low room-temperature mobility measured here ($\mu \approx 1$) suggests that additional extrinsic scattering mechanisms are present.
\begin{wrapfigure}{r}{0.5\textwidth}
    \centering
    \includegraphics[width=0.48\textwidth]{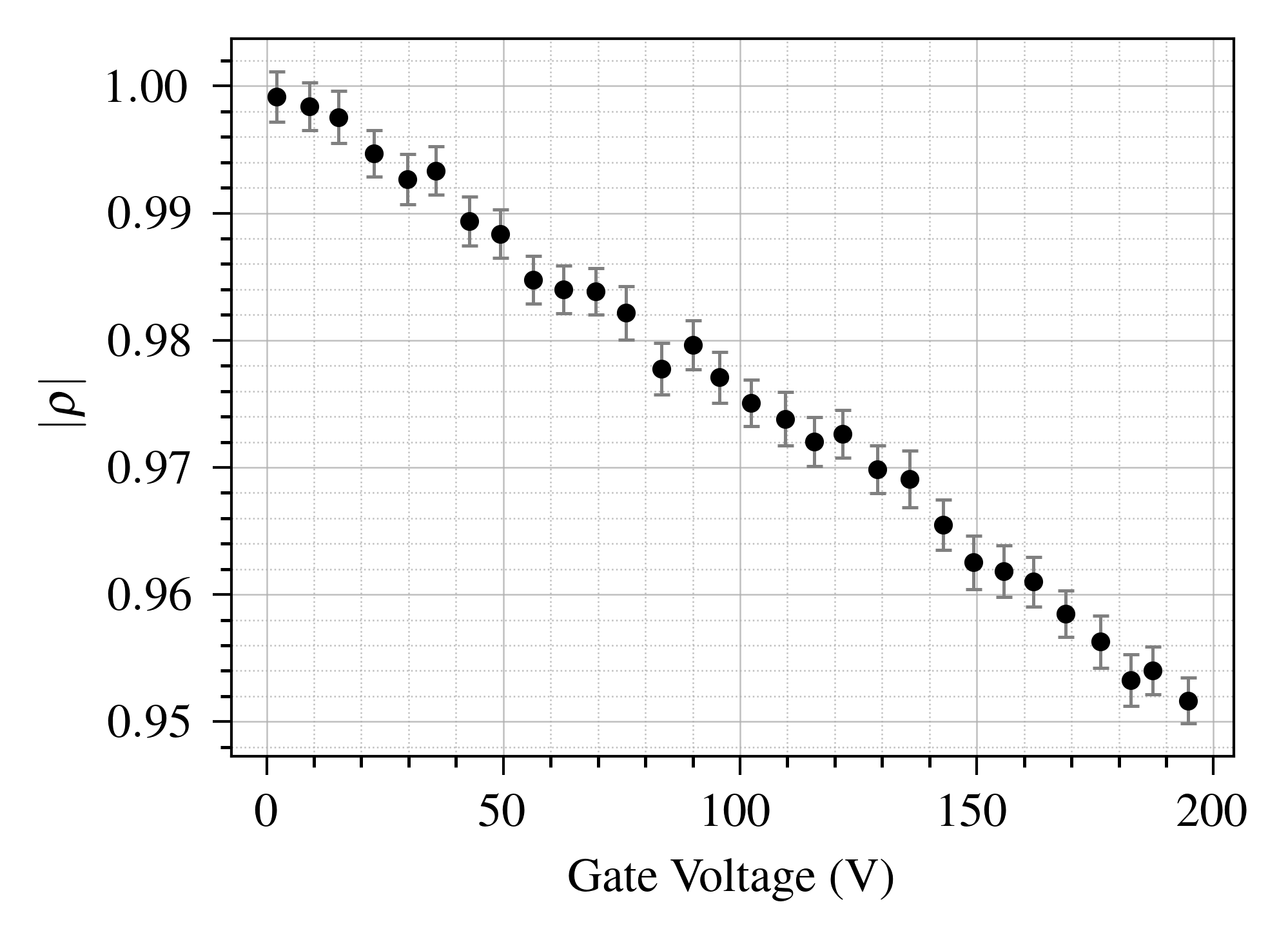}
    \caption{\label{fig:gate_conductivity}
    Gate-dependent resistivity of a single-grain type-II Si clathrate device measured at 300~K. The monotonic decrease in resistivity with increasing gate voltage is consistent with electron accumulation, demonstrating room-temperature electrostatic tunability.}
\end{wrapfigure}
One plausible extrinsic contribution is carrier scattering at crystal-phase boundaries, which is known to strongly suppress and flatten the temperature dependence of $\mu$ in large-grain polysilicon~\cite{Murti1990}. While such boundaries can dominate transport in polycrystalline materials, our single-grain resistivity (Fig.~\ref{fig:rho_vs_T}) and $n(T)$ (Fig.~\ref{HallResults}) retain the expected temperature evolution for a lightly doped semiconductor, arguing against boundary-dominated conduction. We therefore interpret the data as follows: phase boundary scattering likely sets a low ceiling for $\mu$, but the persistence of some temperature dependence suggests that clathrate-specific phonon scattering strongly suppresses carrier mobility relative to similar semiconductors. This is consistent with the results of thermal transport studies: the lattice thermal conductivity of empty Si$_{136}$ is only $\sim$2.5~W\,m$^{-1}$\,K$^{-1}$ at 300 K---roughly thirty times lower than that of diamond-Si~\cite{Nolas2003,Wei2022}. First-principles calculations further reveal reduced acoustic-phonon group velocities and lifetimes-- about half those in diamond-Si~\cite{Cohn1999,Toberer2011}.
\subsection{Gate-dependent Transport}
To test the electrostatic tunability of single-grain clathrates, we performed back-gated conductivity measurements at 300~K (Fig.~5). The Si substrate comprises a 500~nm SiO$_2$ layer and a $\sim$8~$\mu$m KMPR dielectric, so relatively large gate biases are required to achieve appreciable electrostatic coupling through the thick insulating stack. The device exhibits a monotonic decrease in the normalized resistivity with increasing gate voltage, consistent with electron accumulation under positive bias. This response confirms $n$-type conduction in type-II clathrates, as expected for Na donor doping, and demonstrates that carriers in the clathrate framework can be modulated electrostatically at room temperature.

The overall modulation is modest (a few percent over a 200~V sweep), consistent with the large band gap ($\sim$1.9~eV) and low free-carrier density of lightly doped clathrates. Similar weak but monotonic gate responses have been reported in other wide-gap semiconductors and oxide systems, where gating primarily perturbs a limited carrier reservoir rather than producing large changes in conductivity. Importantly, to our knowledge, this constitutes the first demonstration of field-effect control in structurally pure type-II clathrates.

Although the magnitude of modulation is smaller than in conventional semiconductors such as Si or GaAs, these measurements establish a proof-of-concept for electrostatic control in clathrate devices. Greater tunability should be achievable with thinner dielectrics, higher-capacitance gate geometries, and/or controlled doping to increase the baseline carrier density. Even modest room-temperature gate response therefore represents an important step toward integrating clathrates into functional field-effect device architectures.

\section{Conclusion}

We report the first comprehensive low-temperature magnetotransport characterization of individual polycrystalline grains of type-II silicon clathrate. By developing a microfabrication process to isolate and contact single polycrystalline grains, we strongly reduce contributions from amorphous intergranular material that have previously confounded transport measurements on polycrystalline films. This has allowed us to probe the characteristic electronic behavior of the clathrate framework.

Our resistivity data reveal distinct thermally activated transport regimes. A freeze-out activation energy of $\epsilon_1 = 65$~meV is consistent with ionization of Na donors in the clathrate lattice, while a second low-temperature crossover is characterized by an effective energy scale $\epsilon_3 = 0.3$~meV. This unusually small energy scale is compatible with NNH, although contributions from VRH cannot be excluded. Zabrodskii analysis indicates a change in the transport exponent $p$ below 10~K, consistent with a crossover within the hopping regime at substantially lower temperature than previously reported. This shift may be a direct consequence of removing amorphous disorder from the system, highlighting the role of sample purity in governing hopping behavior.

Hall measurements further corroborate this picture, showing a sharp reduction in carrier concentration below $\sim$270~K that aligns with the freeze-out transition; simultaneously, $\mu$ increases upon cooling as phonon scattering weakens, although its absolute magnitude remains suppressed. Further, the collected Hall measurements show that although the mobility trend in type-II Si clathrate mirrors that of crystalline Si, the absolute values are suppressed by nearly three orders of magnitude. This suppression is consistent with the clathrate’s phonon dispersion---including low-velocity acoustic branches and low-lying optical modes---which enhances phonon scattering, and may be further exacerbated by residual crystal-phase boundary scattering.

Finally, gate-dependent conductivity measurements demonstrate that electrostatic modulation of single-grain clathrates is feasible at room temperature. Although the relative change in resistivity is modest, the clear response under positive bias confirms that carriers in the clathrate framework can be tuned via standard gating techniques. This proof-of-concept highlights both the potential and limitations of electrostatic control in lightly doped, wide-gap clathrates, and suggests that future devices may benefit from higher-capacitance gating strategies or controlled doping to enhance responsiveness.

Together, these measurements establish a comprehensive picture of low-temperature transport in low-doped, structurally pure type-II Si clathrate. Our results suggest that electronic conduction is governed by sequential intrinsic, freeze-out, and hopping regimes, highly sensitive to disorder and compensation. These results provide a foundation for incorporating silicon clathrates into future electronic and optoelectronic device architectures.

\section*{Acknowledgements \label{Acknowledgements}}
This work was supported by the National Science Foundation, Grant No. 2114569.

\printbibliography

@article{1,
  title = {Synthesis, Crystal Structure, and Transport Properties of Na22Si136},
  author  = {Beekman, M. and Sebastian, C. P. and Grin, Y. and Nolas, G. S.},
  journal = {J. Electron. Mater.},
  year    = {2009},
  volume  = {38},
  pages   = {1136--1141},
  doi     = {10.1007/s11664-008-0641-y},
}

@article{2,
  title = {Electron paramagnetic resonance study of type-II silicon clathrate with low sodium guest concentration},
  author  = {Schenken, W. K. and Liu, Y. and Krishna, L. and Majid, A. A. A. and Koh, C. A. and Taylor, P. C. and Collins, R. T.},
  journal = {Phys. Rev. B},
  year    = {2020},
  volume  = {101},
  pages   = {245204},
  doi     = {10.1103/PhysRevB.101.245204},
}

@article{6,
  title = {Synthesis of Group IV Clathrates for Photovoltaics},
  author  = {Martinez, A. D. and Krishna, L. and Baranowski, L. L. and Lusk, M. T. and Toberer, E. S. and Tamboli, A. C.},
  journal = {IEEE J. Photovoltaics},
  year    = {2013},
  volume  = {3},
  pages   = {1305--1310},
  doi     = {10.1109/JPHOTOV.2013.2276478},
}

@article{8,
  title = {Wide-band-gap si in open fourfold-coordinated clathrate structures},
  author  = {Adams, G. B. and O'Keeffe, M. and Demkov, A. A. and Sankey, O. F. and Huang, Y.-M.},
  journal = {Phys. Rev. B},
  year    = {1994},
  volume  = {49},
  pages   = {8048},
  doi     = {10.1103/PhysRevB.49.8048},
}

@article{Stefanofski2012,
  title = {Synthesis and Structural Characterization of NaxSi136 ($0 < x \leq 24$) Single Crystals and Low-Temperature Transport of Polycrystalline Specimens},
  author  = {Stefanoski, S. and Malliakas, C. and Kanatzidis, M. and Nolas, G.},
  journal = {Inorg. Chem.},
  year    = {2012},
  volume  = {51},
  pages   = {8686--8692},
  doi     = {10.1021/ic202199t},
}

@article{Yinan2021,
  title = {Synthesis and characterization of type II silicon clathrate films with low Na concentration},
  author  = {Liu, Y. and Schenken, W. K. and Krishna, L. and Majid, A. A. A. and Furtak, T. E. and Walker, M. and Koh, C. A. and Taylor, P. C. and Collins, R. T.},
  journal = {Appl. Phys. Rev.},
  year    = {2021},
  volume  = {8},
  pages   = {041408},
  doi     = {10.1063/5.0062723},
}

@article{13,
  title = {Silicon Clathrate Films for Photovoltaic Applications},
  author  = {Fix, T. and Vollondat, R. and Ameur, A. and Roques, S. and Rehspringer, J. and Chevalier, C. and Muller, D. and Slaoui, A.},
  journal = {J. Phys. Chem. C},
  year    = {2020},
  volume  = {124},
  pages   = {14972--14977},
  doi     = {10.1021/acs.jpcc.0c02712},
}

@article{14,
  title = {Thin film of guest-free type-II silicon clathrate on Si(111) wafer},
  author  = {Kume, T. and Ohashi, F. and Sakai, K. and Fukuyama, A. and Imai, M. and Udono, H. and Ban, T. and Habuchi, H. and Suzuki, H. and Ikari, T. and Sasaki, S. and Nonomura, S.},
  journal = {Thin Solid Films},
  year    = {2016},
  volume  = {609},
  pages   = {30--34},
  doi     = {10.1016/j.tsf.2016.03.056},
}

@article{15,
  title = {Multilayer Fabrication of Micromolding and Electroforming with the Planarization of Grinding for High-Aspect-Ratio Microelectrodes in Electro-conjugate Fluid (ECF) Micropumps},
  author  = {Han, D. and Yamada, Y. and Yokota, S. and Kim, J.-W.},
  journal = {Int. J. Precis. Eng. Manuf.},
  year    = {2020},
  volume  = {21},
  doi     = {10.1007/s12541-019-00299-3},
}

@article{16,
  title = {Variation of thermal and mechanical properties of KMPR due to processing parameters},
  author  = {Reynolds, M. and Elias, A. and Elliott, D. and Backhouse, C. and Sameoto, D.},
  journal = {J. Micromech. Microeng.},
  year    = {2012},
  volume  = {22},
  pages   = {125023},
  doi     = {10.1088/0960-1317/22/12/125023},
}

@article{beekman2015,
  title={Inorganic Clathrate-II Materials: Synthesis, Structure, and Thermoelectric Properties},
  author  = {Beekman, M. and Nolas, G. S.},
  journal = {J. Mater. Chem. C},
  year    = {2015},
  volume  = {3},
  number  = {18},
  pages   = {4692--4708},
  doi = {10.1557/PROC-1044-U05-05}
}

@article{gryko2000,
  title = {Low-density framework form of crystalline silicon with a wide optical band gap},
  author  = {Gryko, J. and McMillan, P. F. and Marzke, R. F. and Ramachandran, G. K. and Patton, D. and Deb, S. K. and Sankey, O. F.},
  journal = {Phys. Rev. B},
  year    = {2000},
  volume  = {62},
  pages   = {R7707--R7710},
  doi = {10.1103/PhysRevB.62.R7707}
}

@book{nolas2014review,
  editor    = {Nolas, G. S.},
  title     = {The Physics and Chemistry of Inorganic Clathrates},
  series    = {Springer Series in Materials Science},
  volume    = {199},
  publisher = {Springer},
  year      = {2014},
  doi = {10.1007/978-94-017-9127-4}
}

@article{shankar2010,
  title={Spin relaxation and coherence times for electrons at the Si/SiO2 interface},
  author  = {Shankar, S. and Tyryshkin, A. M. and He, J. and Lyon, S. A.},
  journal = {Phys. Rev. B},
  year    = {2010},
  volume  = {82},
  pages   = {195323},
  doi = {doi.org/10.1103/PhysRevB.82.195323}
}

@article{kasper1965,
  author  = {Kasper, J. S. and Hagenmuller, P. and Pouchard, M. and Cros, C.},
  journal = {Science},
  year    = {1965},
  volume  = {150},
  pages   = {1713--1714},
  title={Clathrate Structure of Silicon Na8Si46 and NaxSi136 ($x < 11$)},
  doi = {10.1126/science.150.3704.1713}
}

@article{moriguchi2000,
  title = {Electronic structures of ${\mathrm{Na}}_{8}{\mathrm{Si}}_{46}$ and ${\mathrm{Ba}}_{8}{\mathrm{Si}}_{46}$},
  author = {Moriguchi, K. and Yonemura, M. and Shintani, A. and Yamanaka, S.},
  journal = {Phys. Rev. B},
  volume = {61},
  issue = {15},
  pages = {9859--9862},
  year = {2000},
  doi = {10.1103/PhysRevB.61.9859},
}

@article{saito1995,
  title = {Electronic structure of ${\mathrm{Si}}_{46}$ and ${\mathrm{Na}}_{2}$${\mathrm{Ba}}_{6}$${\mathrm{Si}}_{46}$},
  author = {Saito, S. and Oshiyama, A.},
  journal = {Phys. Rev. B},
  volume = {51},
  issue = {4},
  pages = {2628--2631},
  year = {1995},
  doi = {10.1103/PhysRevB.51.2628},
}

@article{kovnir2004,
doi = {10.1070/RC2004v073n09ABEH000916},
year = {2004},
volume = {73},
number = {9},
pages = {923},
author = {Kovnir, K. A. and Shevelkov, A. V.},
title = {Semiconducting clathrates: synthesis, structure and properties},
journal = {Russian Chemical Reviews}
}

@article{Mott1973Clathrates,
  author  = {Mott, N. F. and Davis, E. A.},
  journal = {J. Solid State Chem.},
  title = {Properties of compounds of type Na\(_x\)Si\(_{136}\) and Na\(_8\)Si\(_{46}\)},
  year    = {1973},
  volume  = {6},
  number  = {3},
  pages   = {348--351},
  doi     = {10.1016/0022-4596(73)90221-1},
}

@article{Danilyuk2017,
  title = {Low Temperature Conductivity in n -Type Noncompensated Silicon below Insulator-Metal Transition},
  author  = {Danilyuk, A. and Trafimenko, A. and Fedotov, A. and Svito, I. and Prischepa, S.},
  journal = {Advances in Condensed Matter Physics},
  year    = {2017},
  volume  = {2017},
  pages   = {1--12},
  doi     = {10.1155/2017/5038462},
}

@book{ShklovskiiEfros1984,
  author    = {Shklovskii, B. I. and Efros, A. L.},
  title     = {Electronic Properties of Doped Semiconductors},
  series    = {Springer Series in Solid-State Sciences},
  publisher = {Springer},
  year      = {1984},
}

@article{Poklonski2025,
  title = {Activation Energy of DC Hopping Conductivity of Lightly Doped Weakly Compensated Crystalline Semiconductors},
  author  = {Poklonski, N. A. and Anikeev, I. I. and Vyrko, S. A. and Zabrodskii, A. G.},
  journal = {Phys. Status Solidi B},
  year    = {2025},
  volume  = {262},
  number  = {1},
  pages   = {2400132},
  doi     = {10.1002/pssb.202400132},
}

@article{Miller1960,
  
  author  = {Miller, A. and Abrahams, E.},
  journal = {Phys. Rev.},
  year    = {1960},
  volume  = {120},
  number  = {3},
  pages   = {745--755},
  doi     = {10.1103/PhysRev.120.745},
}

@article{Stefanoski2010,
  title = {Low temperature transport properties and heat capacity of single-crystal Na8Si46},
  author  = {Stefanoski, S. and Martin, J. B. and Nolas, G. S.},
  journal = {J. Phys.: Condens. Matter},
  year    = {2010},
  volume  = {22},
  number  = {48},
  pages   = {485404},
  doi     = {10.1088/0953-8984/22/48/485404},
}

@article{Doubrava1970,
  title        = {The Electrical Properties of Silicon with Sodium Donors},
  author  = {Doubrava, P.},
  journal = {Phys. Status Solidi B},
  year    = {1970},
  volume  = {40},
  number  = {2},
  pages   = {483--492},
  doi     = {10.1002/pssb.19700400205},
}

@article{ZabrodskiiZinovyeva1984,
  title        = {Low-temperature conductivity and the metal--insulator transition in compensated n-type germanium},
  author  = {Zabrodskii, A. G. and Zinovyeva, K. N.},
  journal = {Soviet Physics JETP},
  year    = {1984},
  volume  = {59},
  number  = {2},
  pages   = {425--432},
}

@article{Ambegaokar1971,
  title = {Hopping Conductivity in Disordered Systems},
  author  = {Ambegaokar, V. and Halperin, B. I. and Langer, J. S.},
  journal = {Phys. Rev. B},
  year    = {1971},
  volume  = {4},
  number  = {8},
  pages   = {2612--2620},
  doi     = {10.1103/PhysRevB.4.2612},
}

@article{Castner1999,
  title = {Deconvolution of activated and variable-range-hopping conduction for barely insulating arsenic-doped silicon},
  author  = {Castner, T. G. and Shafarman, W. N.},
  journal = {Phys. Rev. B},
  year    = {1999},
  volume  = {60},
  number  = {20},
  pages   = {14182--14185},
  doi     = {10.1103/PhysRevB.60.14182},
}

@article{Zabrodskii1981,
  title     = {Determination of the conductivity mechanism from the temperature dependence of conductivity},
  author  = {Zabrodskii, A. G. and Zinov'eva, K. N.},
  journal = {Soviet Journal of Low Temperature Physics},
  year    = {1981},
  volume  = {7},
  pages   = {331--334},
}

@article{Ray1961,
  title        = {Impurity Conduction in Silicon},
  author  = {Ray, R. K. and Fan, H. Y.},
  journal = {Phys. Rev.},
  year    = {1961},
  volume  = {121},
  number  = {3},
  pages   = {768--776},
  doi     = {10.1103/PhysRev.121.768},
}

@article{Cros1970,
  title        = {Preparation and Characterization of the Silicon Clathrate NaxSi\textsubscript{136} ($x \leq 3$): Electrical Transport in the Dilute Sodium Limit},
  author  = {Cros, C. and Pouchard, M. and Hagenmuller, P.},
  journal = {J. Solid State Chem.},
  year    = {1970},
  volume  = {2},
  pages   = {570--581},
  doi     = {10.1016/0022-4596(70)90053-8},
}

@book{Mott1979,
  title     = {Electronic Processes in Non-Crystalline Materials},
  author    = {Mott, N. F. and Davis, E. A.},
  publisher = {Clarendon Press},
  address   = {Oxford},
  edition   = {2nd},
  year      = {1979},
  doi = {10.1088/0031-9112/24/5/033}
}

@article{Stefanoski2011Simple,
  title   = {Simple approach for selective crystal growth of intermetallic clathrates},
  author  = {Stefanoski, S. and Beekman, M. and Wong-Ng, W. and Zavalij, P. Y. and Nolas, G. S.},
  journal = {Chem. Mater.},
  year    = {2011},
  volume  = {23},
  pages   = {1491--1495},
  doi     = {10.1021/cm103135k},
}

@article{Slingsby2016NaDiffusion,
  title   = {Dynamic free energy surfaces for sodium diffusion in type II silicon clathrates},
  author  = {Slingsby, J. G. and Rorrer, N. A. and Krishna, L. and Toberer, E. S. and Koh, C. A. and Maupin, C. M.},
  journal = {Phys. Chem. Chem. Phys.},
  year    = {2016},
  volume  = {18},
  pages   = {5121--5128},
  doi     = {10.1039/C5CP06066D},
}

@article{Briggs2024APL,
  title   = {Characterization of low sodium type II silicon clathrate film spin dynamics},
  author  = {Briggs, J. P. and Liu, Y. and Taylor, P. C. and Singh, M. and Collins, R. T. and Koh, C. A.},
  journal = {Appl. Phys. Lett.},
  year    = {2024},
  volume  = {125},
  number  = {20},
  pages   = {202402},
  doi     = {10.1063/5.0230407},
}

@article{Liu2025,
  title = {Advancements in low-density crystalline silicon allotropes},
  author  = {Liu, Y. and Briggs, J. P. and Collins, R. T. and Singh, M. and Taylor, P. C. and Koh, C. A.},
  journal = {Appl. Phys. Lett.},
  year    = {2025},
  volume  = {126},
  number  = {9},
  pages   = {090501},
  doi     = {10.1063/5.0252405},
}

@phdthesis{vanHapert2002,
    author = {{van Hapert}, J. J.},
    title = "{Hopping Conduction and Chemical Structure : a study on Silicon Suboxides}",
    school = {University of Utrecht, Netherlands},
    year = 2002,
}

@article{Dong1999,
  title     = {Theoretical study of the vibrational modes and their pressure dependence in the pure clathrate-II silicon framework},
  author  = {Dong, J. and Sankey, O. F. and Kern, G.},
  journal = {Phys. Rev. B},
  year    = {1999},
  volume  = {60},
  number  = {2},
  pages   = {950--958},
  doi     = {10.1103/PhysRevB.60.950},
}

@article{Wei2022,
  title     = {Structural Dynamics, Phonon Spectra and Thermal Transport in the Silicon Clathrates},
  author  = {Wei, B. and Flitcroft, J. M. and Skelton, J. M.},
  journal = {Molecules},
  year    = {2022},
  volume  = {27},
  number  = {19},
  pages   = {6431},
  doi     = {10.3390/molecules27196431},
}

@article{Nolas2003,
  title     = {Thermal conductivity of elemental crystalline silicon clathrate {Si$_{136}$}},
  author  = {Nolas, G. S. and Beekman, M. and Gryko, J. and Lamberton, J. R. and Tritt, T. M. and McMillan, P. F.},
  journal = {Appl. Phys. Lett.},
  year    = {2003},
  volume  = {82},
  number  = {6},
  pages   = {910--912},
  doi     = {10.1063/1.1544640},
}

@article{Cohn1999,
  title     = {Glasslike Heat Conduction in High-Mobility Crystalline Semiconductors},
  author  = {Cohn, J. L. and Nolas, G. S. and Fessatidis, V. and Metcalf, T. H. and Slack, G. A.},
  journal = {Phys. Rev. Lett.},
  year    = {1999},
  volume  = {82},
  number  = {4},
  pages   = {779--782},
  doi     = {10.1103/PhysRevLett.82.779},
}

@article{Toberer2011,
  title     = {Phonon engineering through crystal chemistry},
  author  = {Toberer, E. S. and Zevalkink, A. and Snyder, G. J.},
  journal = {J. Mater. Chem.},
  year    = {2011},
  volume  = {21},
  number  = {40},
  pages   = {15843--15852},
  doi     = {10.1039/C1JM11754H},
}

@article{Morin1954,
  title   = {Electrical Properties of Silicon Containing Arsenic and Boron},
  author  = {Morin, F. J. and Maita, J. P.},
  journal = {Phys. Rev.},
  year    = {1954},
  volume  = {96},
  number  = {1},
  pages   = {28--35},
  doi     = {10.1103/PhysRev.96.28},
}

@article{Friedman1981,
  title   = {The Hall effect in the variable-range-hopping regime},
  author  = {Friedman, L. and Pollak, M.},
  journal = {Philos. Mag. B},
  year    = {1981},
  volume  = {44},
  number  = {4},
  pages   = {487--507},
  doi     = {10.1080/01418638108222584},
}

@article{Avdonin2016,
  author  = {Avdonin, A. and Skupi{\'n}ski, P. and Grasza, K.},
  title   = {Hall effect in hopping regime},
  journal = {Physica B},
  year    = {2016},
  volume  = {483},
  pages   = {13--18},
  doi     = {10.1016/j.physb.2015.12.024},
}

@article{Murti1990,
  author  = {Murti, M. R. and Reddy, K. V.},
  title = {Grain Boundary Effects on the Carrier Mobility of Polysilicon},
  journal = {Phys. Status Solidi A},
  year    = {1990},
  volume  = {119},
  number  = {1},
  pages   = {237--240},
  doi     = {10.1002/pssa.2211190128},
}
\end{document}